\documentclass[superscriptaddress,twocolumn,nofootinbib,showpacs,showkeys,pra,amsmath]{revtex4}

\usepackage{graphicx}
\usepackage{xcolor,hyperref}

\newcommand{\ket}[1]{|#1\rangle}
\newcommand{\bra}[1]{\langle#1|}

\newcommand{\eq}{\begin{equation}}
\newcommand{\fine}{\end{equation}}

\newcommand{\kk}{{\bf k}}
\newcommand{\cc}{{\bf c}}

\begin{document}


\title{Hyperentanglement of two photons in three degrees of freedom}
\author{Giuseppe Vallone}
\homepage{http://quantumoptics.phys.uniroma1.it/}
\affiliation{
Dipartimento di Fisica della Sapienza Universit\`{a} di Roma,
Roma, 00185 Italy and\\
Consorzio Nazionale Interuniversitario per le Scienze Fisiche della Materia,
Roma, 00185 Italy}
\author{Raino Ceccarelli}
\homepage{http://quantumoptics.phys.uniroma1.it/}
\affiliation{
Dipartimento di Fisica della Sapienza Universit\`{a} di Roma,
Roma, 00185 Italy and\\
Consorzio Nazionale Interuniversitario per le Scienze Fisiche della Materia,
Roma, 00185 Italy}
\author{Francesco De Martini}
\affiliation{
Dipartimento di Fisica della Sapienza Universit\`{a} di Roma,
Roma, 00185 Italy and\\
Consorzio Nazionale Interuniversitario per le Scienze Fisiche della Materia,
Roma, 00185 Italy}
\affiliation{Accademia Nazionale dei Lincei, via della Lungara 10, Roma 00165, Italy}
\homepage{http://quantumoptics.phys.uniroma1.it/}
\author{Paolo Mataloni}
\homepage{http://quantumoptics.phys.uniroma1.it/}
\affiliation{
Dipartimento di Fisica della Sapienza Universit\`{a} di Roma,
Roma, 00185 Italy and\\
Consorzio Nazionale Interuniversitario per le Scienze Fisiche della Materia,
Roma, 00185 Italy}

\date{\today}

\begin{abstract}
A $6$-qubit hyperentangled state has been realized by entangling two photons in three degrees of freedom. 
These correspond to the polarization, the longitudinal momentum and the indistinguishable emission produced by a $2$-crystal system 
operating with Type I phase matching in the spontaneous parametric down conversion regime. 
The state has been characterized by a chained interferometric apparatus and its complete entangled nature has been 
tested by a novel witness criterium specifically introduced for hyperentangled states. 
The experiment represents the first realization of a genuine hyperentangled state with the maximum entanglement 
between the two particles allowed in the given Hilbert space.
\end{abstract}

\pacs{03.67.Bg, 42.50.Dv, 42.65.Lm}
\keywords{hyperentanglement, spontaneous parametric down conversion, multiqubit photon states}
\maketitle

Hyperentangled (HE) $2$-photon states, i.e. quantum states 
of two photons simultaneously entangled in $N$ independent 
degrees of freedom (DOF's) \cite{kwia97jmo, barb05pra}, are nowadays 
recognized as a basic resource for many quantum information (QI) 
processes.
At variance with $n$-photon entanglement, hyperentanglement is much 
less affected by decoherence and the detection efficiency of the state scales by $\eta^{n}$,  
with $\eta$ the quantum efficiency of detectors,
is not reduced by growing the size of the state.
 
Important quantum communication protocols have been demonstrated by using
two photon HE states, such as enhancing the 
channel capacity in superdense coding \cite{barr08nat} and the complete and 
deterministic discrimination of the four orthogonal polarization Bell 
states \cite{schu06prl, barb07pra}.
Hyperentanglement is also a viable resource in view of enhancing 
the power of computation of a scalable one-way quantum 
computer \cite{brie01prl}. Indeed, by applying suitable 
transformations on the HE states, efficient $4$-qubit $2$-photon cluster 
states were created \cite{vall07prl} and used in the realization of 
basic quantum computation algorithms \cite{chen07prl, vall08prl}. 
Finally, working in a large dimension Hilbert space with $N$ entangled DOF's of the photons allows 
to perform advanced tests of quantum nonlocality able to largely 
enhance the violation of local realism 
\cite{coll02prl, barb06prl, cabe06prl}.
\begin{figure*}[t]
\centering
 \includegraphics[width=12cm]{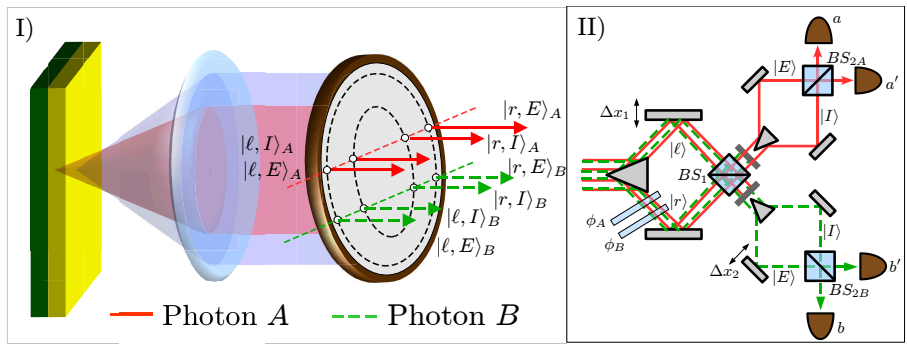}
\caption{Experimental setup. I) The $2$-crystal system generates photon pairs with equal probability 
over the internal ($I$) and external ($E$) cone. Polarization entanglement arises 
from double passage of the pump beam through the crystals and the combined action of a 
spherical mirror and a $\lambda/4$ wave-plate (not shown in the figure). The annular sections
of each emission cone, with diameters $d_I=12mm$ and $d_E=16mm$ are intercepted by a single $8$-hole screen.
II) Measurement setup: the first interferometer (ending with $BS_1$) measures momentum entanglement in the basis $\frac1{\sqrt2}(\ket\ell+e^{i\phi_{A,B}}\ket r)$, 
at the same time for Alice (red line) and Bob (dotted green line) photon. Phase $\phi_A$ ($\phi_B$) is set by tilting a thin
glass plate intercepting the Alice (Bob) mode.
The subsequent interferometers, corresponding to $BS_{2A}$ and $BS_{2B}$, perform the measurement
separately for the Alice and Bob photon in the basis $\frac1{\sqrt2}(\ket I\pm\ket E)$.
The interference filters allowing photon detection within a bandwidth $\Delta\lambda = 6nm$
and the four apparatus for polarization analysis aligned before each detector are not shown
in the figure.}
\label{fig:source}
\end{figure*}

HE states of increasing size are of paramount importance for the realization of 
even more challenging QI objectives. Recently, optical schemes involving
two or more photons were proposed to expand the available Hilbert space to
incredibly large dimensions.
In particular three DOF's of the photons, namely polarization, spatial mode, 
and time-energy, were used to build a multidimension entangled state of two photons
encoded in four qubits and two qutrits \cite{barr05prl}. More recently, Greenberger-Horne-Zeilinger 
(GHZ) states of up to $10$-qubits were engineered by exploiting polarization 
and momentum of five photons \cite{gao08qph}.
It is worth noting that in both experiments the addition of one DOF, i.e.
time-energy and linear momentum, respectively, derives from a local 
manipulation\footnote{We define local manipulation any transformation
that acts separately on each particle.}
of polarization entanglement that enhances the number of DOF's without increasing 
the entanglement between the particles.
Indeed, in both experiments the entanglement of the additonal DOF was not independent 
from the polarization entanglement\footnote{In the experiment of Ref. 
\cite{barr05prl} two Franson-type polarization
interferometers, one for each photon and each one realized by a quartz birefringent element, 
were used to analyze time-energy entanglement. The entanglement due to the polarization 
and time-energy DOF's could be made really ``independent'' and thus the total entanglement 
of the state could grow by using standard 
Franson-type unbalanced interferometers \cite{fran89prl}.}.

We can fully take advantage of an enlarged Hilbert space by independently entangling the
two photons in $N$ DOF's in order to create a ``genuine'' HE state,
given by the product of $N$ Bell states, 
\eq
\ket{\Psi_N} = \ket{\text{Bell}_1}\otimes\ket{\text{Bell}_2} \otimes\cdots\otimes\ket{\text{Bell}_N}
\fine
one for each DOF \cite{vall08qph}.

Let's consider the {\it entropy of entanglement} of the state $\ket{\Psi_N}$,
\eq
E(\ket{\Psi_N}) = S[\text{tr}_A(\ket{\Psi_N}\bra{\Psi_N})] = S[\text{tr}_B(\ket{\Psi_N}\bra{\Psi_N})]\,,
\fine
where $S$ is the standard Von-Neumann entropy, defined as $S(\rho)=\text{tr}(\rho\log_2\rho)$.
It's easy to demonstrate that $E(\ket\Psi) = N$,  independently of any
local operation. Hence the state $\ket\Psi$ allows to maximize the degree of 
entanglement of two particles for a given Hilbert space. On the other hand, by adding 
other DOF's to the state by local operations, the entropy of entanglement doesn't change, 
as said.

In this Letter we present the experimental realization and characterization 
of a $6$-qubit HE state based on the simultaneous triple entanglement
of two photons.
By our scheme, besides polarization ($\pi$) entanglement, two photons experience 
a further double entanglement in the longitudinal momentum DOF. In the experiment this was done by 
selecting four pairs of correlated {\bf k}-modes within the spatial emission of a 
two-crystal system operating in the spontaneous parametric down conversion (SPDC) 
regime. 
This configuration realizes a genuine $2$-photon HE state with $E=3$ and thus entangled in $N = 3$ 
independent DOF's, labelled as $\pi$, $\kk$ and $\cc$.

In the experiment we adopted a $2$-crystal geometry to generate the $6$-qubit HE 
state. Precisely, two 0.5 mm thick Type I $\beta$-barium-borate (BBO) crystal slabs, 
cut at different phase matching angles, (see Figure \ref{fig:source}-I), aligned one behind the other, 
were operating under double excitation (back and forth) of a vertically ($V$) polarized 
continuous wave (CW) laser at wavelength $\lambda_p$. 

The polarization entangled Bell state 
$\ket{\Phi^-}=\frac{1}{\sqrt{2}}\left( \ket H_A\ket H_B-\ket V_A\ket V_B\right) $ 
was created by spatial superposing the two perpendicularly polarized emission
cones of each Type I crystal, in agreement with a method already described 
by us in previous papers \cite{barb05pra}.
\begin{figure*}[t]
\centering
 \includegraphics[width=7.5cm]{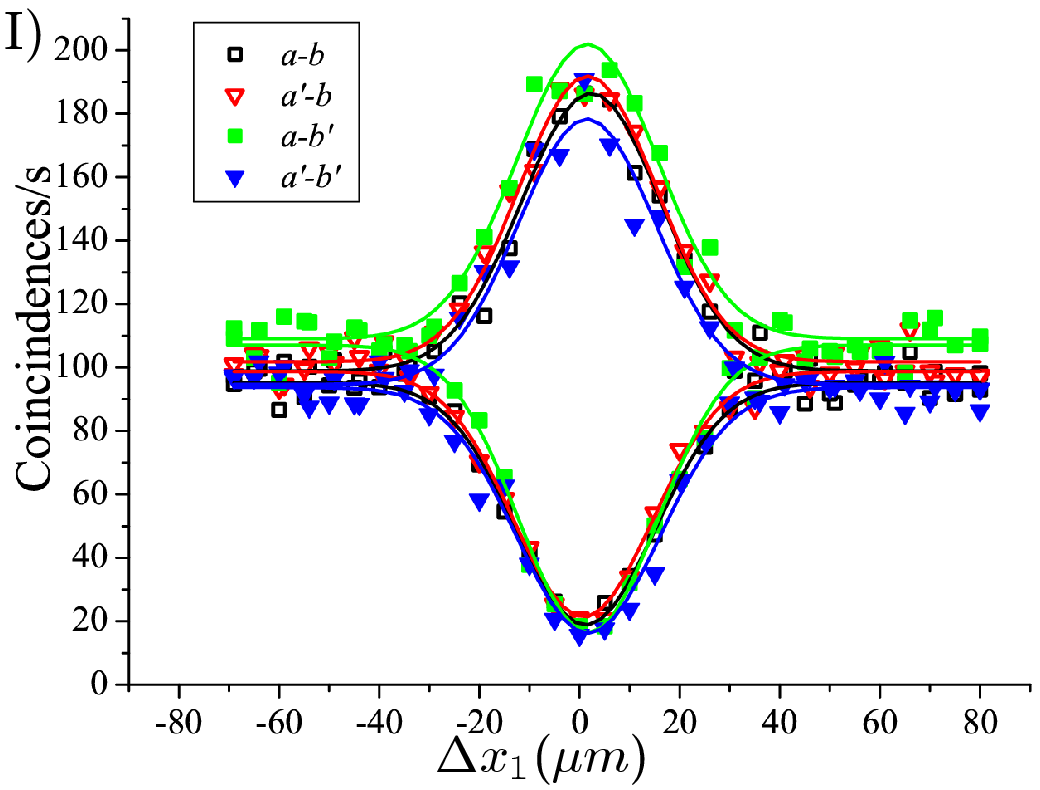}\quad\quad
 \includegraphics[width=7.5cm]{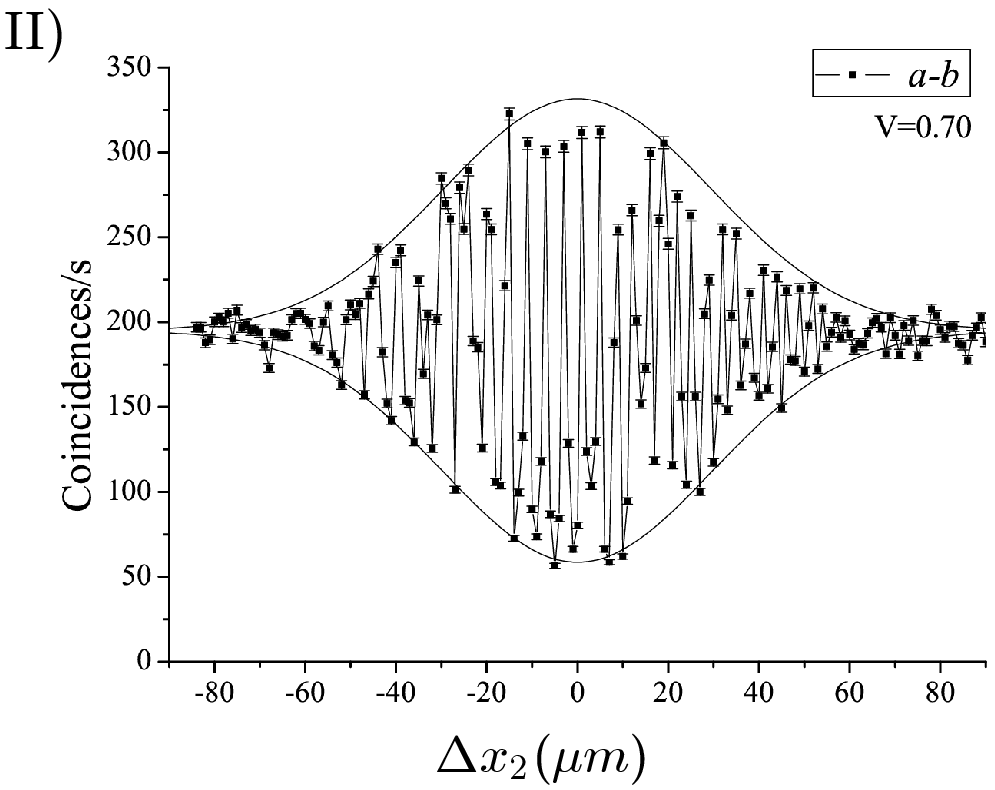}
\caption{I) Coincidence rates $a$-$b$, $a'$-$b$, $a$-$b'$, $a'$-$b'$ vs $\Delta x_1$ for the
state \eqref{HE} showing the typical dip- ($\varphi_\kk = 0$) peak ($\varphi_\kk = \pi$) feature 
with average visibility $V = 0.815\pm0.022$ (dip) and $V = 0.877\pm0.031$ (peak). 
The corresponding experimental fits are also shown. 
FWHM $\simeq 30 nm$ of the interference patterns is in agreement
with the expected value for a filter bandwidth of 6 nm. II) Coincidence rate $a$-$b$ vs $\Delta x_2$ for the
state \eqref{HE} with $\varphi_\kk=\pi$ with visibility $V \simeq 0.70$ and
FWHM $\simeq= 60$ nm. In both cases the time average of each experimental point is $3 s$.}
\label{fig:interf}
\end{figure*}

Photon pairs were created with equal probability at degenerate wavelength $\lambda=2\lambda_p$ 
by either one of BBO crystals along two correlated directions belonging to the lateral 
surfaces of two SPDC cones, with full aperture angles $\theta_I=4.6^\circ$ and $\theta_E=6.1^\circ$,
respectively. We refer to them as the ``internal'' ($I$) and the ``external'' ($E$) 
cone, corresponding to the first and the second crystal, respectively.
The dichotomy existing between the $I$ and $E$ cone ``photon choice'' is thus 
identified as a third independent DOF in the state.
Coherence and hence indistinguishability between the two crystal emissions
is guaranteed by the the fact that the pump coherence length exceeds by more 
than one order of magnitude the total crystal length. 

The two conical emissions were then transformed into two cylindrical ones by a positive lens
with focal length $f = 15 cm$, located at a distance $f$ from the intermediate point of the $2$-crystal device. 
Double longitudinal momentum entanglement was created by selecting four pairs of correlated 
modes with a $8$-hole screen. Precisely, the following correlated mode pairs, 
$\ket{I,\ell}_A$ - $\ket{I,r}_B$, $\ket{I,r}_A$ - $\ket{I,\ell}_B$ were selected for the first crystal 
and $\ket{E,\ell}_A$ - $\ket{E,r}_B$, $\ket{E,r}_A$ - $\ket{E,\ell}_B$ for the second crystal emission.
It is worth to remember that only two photons, namely A (Alice) and B (Bob), are generated with equal probability by 
either one of the crystals. We label the corresponding mode emission as $\ell$ ($r$) by referring to the left 
(right) side of each cone and as $I$ ($E$) by considering the internal (external) cone (see Figure \ref{fig:source}-I)).

The produced state is given by the product 
of one polarization entangled state and two longitudinal momentum entangled states (or,
equivalently, two ququart entangled state) and is expressed as a $6$-qubit HE state: 
\begin{equation}\label{HE}
\begin{aligned}
&\frac{1}{\sqrt2}\left(\ket{H}_A\ket{H}_B+e^{i\varphi_\pi}\ket{V}_A\ket{V}_B\right)\otimes
\frac{1}{\sqrt{2}}\left( \ket{\ell}_A\ket{r}_B\right.
\\
&\left.+e^{i\varphi_\kk}\ket{r}_A\ket{\ell}_B\right) \otimes
\frac{1}{\sqrt{2}}
\left( \ket{I}_A\ket{I}_B+e^{i\varphi_\cc}\ket{E}_A\ket{E}_B\right)
\end{aligned}%
\end{equation}

It's easy to verify that the entropy of entanglement of this state equals exactly the 
number of DOF's $N=3$. It corresponds to the maximum allowed 
entanglement for a $2^3\times 2^3$ Hilbert space, as said.

The measurement setup used to test the triple entanglement was given 
by the chained interferometric apparatus sketched in Figure \ref{fig:source}-II). 
It is divided in two parts: the first one performs the measurement of the momentum entanglement existing within each 
emission cone. Here the $\ell$ modes are spatially and temporally matched with the 
$r$ modes both for the up- ($A$ photon) and down- ($B$ photon) side of a common 
50:50 beam splitter ($BS_1$), for both the $I$ and $E$ emission cone.     
In the second part of the apparatus, measuring momentum entanglement between the two  
crystal emissions, the left (right) $BS_1$ output modes belonging to the Alice (Bob) 
photon are then injected in another interferometer where the $I$ and $E$ modes are 
matched onto the $BS_{2A}$ ($BS_{2B}$) beam splitter.

Two trombone mirror assemblies mounted on motorized translation stages allow
fine adjustment of path delays $\Delta x_1$ and $\Delta x_2$ between the different mode sets in 
the first and second part of the interferometric apparatus. 
Phase setting of $\kk$ momentum entanglement is achieved by careful tiliting of thin glass 
plates $\phi_A$ and $\phi_B$, placed respectively on the Alice and Bob right modes (cfr. Fig. \ref{fig:source}-II).

Four single photon detectors, mod. Perkin Elmer SPCM-AQR14, detect 
the radiation belonging to the Alice output modes of $BS_{2A}$, namely $a$ and $a'$ in Figure \ref{fig:source}-II),
and to the Bob output modes of $BS_{2B}$ ($b$ and $b'$ in Figure \ref{fig:source}-II)).
Polarization entanglement is measured by using four standard polarization analyzers,
one for each detector.  

The experimental results given in Figure \ref{fig:interf}-I) show the characteristic interference
patterns obtained by measuring the coincidences $a-b$, $a-b'$, $a'-b$ and $a'-b'$ as a function
of $\Delta x_1$. Momentum entanglement is demonstrated by the typical dip-peak transition 
obtained by adjusting the state phase.
In the same way, similar interference traces were obtained for the four sets of coincidences 
by varying the delay $\Delta x_2$ in the second interferometer. We show in Figure \ref{fig:interf}-II) the one
corresponding to the measurement of $a-b$ coincidences.

We can observe the different effect on phase stability appearing in the figures:
in the first interferometer ($\ket\ell$-$\ket r$ measurement) the phase $\varphi_\kk$ is self-stabilized 
since any mirror instability corresponds to a global phase variation of the state.
Moreover, since any change of $\Delta x_1$ determines the simultaneous variation of all the $\ket r$ 
mode optical delays, even in this case the phase remains constant. 
This features is evident in Fig. \ref{fig:interf}-I),
where we show the different coincidence patterns for $\varphi_\kk=0$ and $\varphi_\kk=\pi$.

In the second interferometer ($\ket I$-$\ket E$ measurement) the phase $\varphi_\cc$ is not self-stabilized
and a compact interferometer is needed to achieve high phase stability.
Moreover, the variation of $\Delta x_2$ changes only the $\ket E$ path delay of the $B$ photon. This corresponds
in Fig. \ref{fig:interf}-II) to a random $\varphi_\cc$ phase variation.
For the same reason the full width haf maximum (FWHM) duration of the interference 
pattern of Fig. \ref{fig:interf}-II) doubles the one of Figure \ref{fig:interf}-I).

We detected the entanglement present in the generated HE state by using a novel 
criterium recently proposed for hyperentanglement \cite{vall08qph}.
It is based on the evaluation of some witnesses specifically introduced for graph states.

Let's describe how the hyperentanglement test was carried out. 
For each DOF spanning a two dimensional Hilbert space, we associate the 
states $\ket H$, $\ket \ell$ and $\ket I$ to the basis vector $\ket0$,
while $\ket V$, $\ket r$ and $\ket E$ are associated to $\ket 1$.
By setting all the phases $\varphi_\mu$ (with $\mu=\pi$, $\kk$, $\cc$) to zero we obtain the following HE state:
\eq\label{Xi}
\ket{\Psi_3}=\ket{\phi^+}_\pi\otimes\ket{\psi^+}_\kk\otimes\ket{\phi^+}_\cc
\fine
with the standard Bell states $\ket{\phi^+}=\frac1{\sqrt2}(\ket0_A\ket0_B+\ket0_A\ket0_B)$
and $\ket{\psi^+}=\frac1{\sqrt2}(\ket0_A\ket1_B+\ket1_A\ket0_B)$.
We also define $Z_i$, $z_i$ and $z'_i$ ($i=A,B$) the $\sigma_z$ Pauli matrices for the $\pi$, $\kk$ and $\cc$ DOF respectively.
Similarly, $X_i$, $x_i$ and $x'_i$ represent the corresponding $\sigma_x$ matrices.

We first verified the presence of entanglement between the two photons for each separate DOF.
This was done by tracing the other DOF's and evaluating separately an entanglement witness for 
$\pi$, $\kk$ and $\cc$ entanglement.
Let's define the following witnesses:
\begin{subequations}
\begin{align}
W_\pi&=(\openone_\pi-2X_AX_B-2Z_AZ_B)\otimes\openone_\kk\otimes\openone_\cc\\
W_\kk&=\openone_\pi\otimes(\openone_\kk-2x_Ax_B-2z_Az_B)\otimes\openone_\cc\\
W_\cc&=\openone_\pi\otimes\openone_\kk\otimes(\openone_\cc-2x'_Ax'_B-2z'_Az'_B)
\end{align}
\end{subequations}

The negative experimental values, $W_\pi=-0.6298\pm0.0080$, $W_\kk=-0.7987\pm0.0055$ 
and $W_\cc=-0.4101\pm0.0082$ demonstrate the entanglement in each DOF.

\begin{ruledtabular}
\begin{table}[t]
\begin{tabular}{ccccccc}
$S_1$           & $S_2$           & $S_3$           & $S_4$           \\
$0.733\pm0.006$ & $0.897\pm0.005$ & $0.810\pm0.005$ & $0.989\pm0.002$ \\
$S_5$           & $S_6$           & $S_1S_3$        & $S_1S_5$        \\
$0.420\pm0.008$ & $0.990\pm0.002$ & $0.681\pm0.007$ & $0.443\pm0.008$ \\
$S_3S_5$        & $S_1S_3S_5$     & $S_2S_4S_6$     & $S_2S_4$        \\
$0.398\pm0.008$ & $0.445\pm0.008$ & $0.891\pm0.005$ & $0.893\pm0.005$ \\
$S_2S_6$        & $S_4S_6$ \\
$0.895\pm0.005$ & $0.988\pm0.002$ 
\end{tabular}
\caption{Experimental values of the operators needed to measure $W_\pi$, $W_\cc$, $W_\kk$, $W_2$, and $W_3$.}
\label{table:operators}
\end{table}
\end{ruledtabular}

As explained in \cite{vall08qph}, these measurement are not sufficient to ensure that the prepared state 
is a genuine HE state. Indeed there are states with negative values of $W_\pi$, $W_\kk$ and $W_\cc$
that can be prepared by a classical mixture of several simple entangled (i.e. entangled in only one DOF) states.
We then need to measure a witness able to detect the global entanglement of the HE state. 
It has been demonstrated \cite{vall08qph}
that the following operators can be used in our case:

\eq
W_2=3-2\left(\prod^3_{k=1}\frac{S_{2k}+1}{2}+\prod^3_{k=1}\frac{S_{2k-1}+1}{2}\right)
\fine
\eq
W_3=2-3\prod^3_{k=1}\left(\frac{1+S_{2k-1}+S_{2k}}{3}\right)
\fine
where
$S_1=X_AX_B$, $S_2=Z_AZ_B$,
$S_3=x_Ax_B$, $S_4=-z_Az_B$,
$S_5=x'_Ax'_B$ and $S_6=z'_Az'_B$ 
are the {\it stabilizers} of the hyperentangled state \eqref{Xi}.
The expectation value of both the witnesses is equal to -1 for a pure triple HE state.  
The difference between $W_2$ and $W_3$ resides on the respective numbers of
measurement settings and on their resistance to white noise.
In particular $W_3$ is more demanding than $W_2$ since it requires a larger number of 
settings but, at the same time, it is more resistant to noise.
The measured values of the witnesses, $W_2=-0.1182\pm0.0055$ and $W_3=-0.0890\pm0.0037$, 
obtained by the experimental values of the operators shown in Table I, demonstrate that 
the whole produced state is entangled and thus represents a genuine HE state. 

In conclusion, we experimentally demonstrated the simultaneous entanglement
of two photons by using three independent DOF's, corresponding to the polarization and 
two different longitudinal momentum deriving by the indistinguishable emission of two nonlinear crystals.
The existence of entanglement of the state was demonstrated by using an entanglement witness method 
specifically introduced for HE state. 
Our results represent the first realization of a $2$-photon $6$-qubit HE state 
with the maximum achievable entanglement for a state involving three DOF's. 
Since the number of $\mathbf{k}$-modes involved in Type I emission processes is 
virtually infinite, HE states with even larger size are expected by using 
longitudinal momentum entanglement but at the price of a large number of 
$\mathbf{k}$-modes in which the photons can be detected.
Other DOF's can be exploited in the future, such as energy-time \cite{fran89prl,ross08pra} 
or orbital angular momentum \cite{barr05prl}, to create larger multidimension entangled 
states of two photons suitable for novel advanced QI tasks.

We thank A. Cabello for useful discussions.


\end{document}